%% file: main.tex
\documentclass[a4paper,11pt]{article}
\pdfoutput=1 

\usepackage{jcappub} 

\usepackage[T1]{fontenc} 
\usepackage{longtable}

\title{\boldmath Review of the online analyses of multi-messenger alerts and electromagnetic transient events with the ANTARES neutrino telescope}

\input{authors}

\abstract{By constantly monitoring a very large portion of the sky, neutrino telescopes are \color{blue}well-designed \color{black} to detect neutrinos emitted by transient astrophysical events. Real-time searches with the ANTARES telescope have been performed to look for neutrino candidates coincident with gamma-ray bursts detected by the \textit{Swift} and \textit{Fermi} satellites, high-energy neutrino events registered by IceCube, transient events from blazars monitored by HAWC, photon-neutrino coincidences by AMON notices and gravitational wave candidates observed by LIGO/Virgo. By requiring temporal coincidence, this approach increases the sensitivity and the significance of a potential discovery. This paper summarises the results of the follow-up performed \color{blue}of \color{black} the ANTARES telescope between January 2014 and February 2022, which corresponds to the end of the \color{blue}data-taking \color{black} period.}

\begin{document}
\maketitle
\flushbottom

\section{Introduction}
\input{introduction.tex}

\section{Follow-up of IceCube neutrino alerts}
\input{IC_follow-up.tex}

\section{Follow-up of LIGO/Virgo gravitational wave alerts}
\input{GW_follow-up.tex}

\section{Follow-up of gamma-ray bursts}
\input{GRB_follow-up.tex}

\section{Follow-up of HAWC alerts for transient phenomena}
\input{transient_follow-up.tex}

\section{Conclusions}
\input{conclusion.tex}

\acknowledgments{The authors acknowledge the financial support of the funding agencies:
Centre National de la Recherche Scientifique (CNRS), Commissariat \`a
l'\'ener\-gie atomique et aux \'energies alternatives (CEA),
Commission Europ\'eenne (FEDER fund and Marie Curie Program), LabEx UnivEarthS (ANR-10-LABX-0023 and ANR-18-IDEX-0001),
R\'egion
Alsace (contrat CPER), R\'egion Provence-Alpes-C\^ote d'Azur,
D\'e\-par\-tement du Var and Ville de La
Seyne-sur-Mer, France;
Bundesministerium f\"ur Bildung und Forschung
(BMBF), Germany; 
Istituto Nazionale di Fisica Nucleare (INFN), Italy;
Nederlandse organisatie voor Wetenschappelijk Onderzoek (NWO), the Netherlands;
Executive Unit for Financing Higher Education, Research, Development and Innovation (UEFISCDI), Romania;
Ministerio de Ciencia e Innovaci\'{o}n: 
Programa Estatal para Impulsar la Investigaci\'{o}n Cient\'{i}fico-T\'{e}cnica
y su Transferencia (refs. PID2021-124591NB-C41, -C42, -C43) (MCIU/FEDER),
Programa de Planes Complementarios I+D+I (refs. ASFAE/2022/023, ASFAE/2022/014)
and Programa Mar\'{i}a Zambrano (Spanish Ministry of Universities, funded by the European Union, NextGenerationEU),
Generalitat Valenciana: Prometeo (PROMETEO/2020/019), 
and GenT (refs. CIDEGENT/2018/034, /2019/043, /2020/049. /2021/23) programs,
Junta de Andaluc\'{i}a (ref. SOMM17/6104/UGR, P18-FR-5057), EU: MSC program (ref. 101025085), Spain;
Ministry of Higher Education, Scientific Research and Innovation, Morocco, and the Arab Fund for Economic and Social Development, Kuwait.
We also acknowledge the technical support of Ifremer, AIM and Foselev Marine
for the sea operation and the CC-IN2P3 for the computing facilities.}

\bibliographystyle{ieeetr}
\bibliography{references}


\end{document}

%% file: authors.tex
\author[1,2]{A.~Albert}
\author[3]{S.~Alves}
\author[4]{M.~Andr\'e}
\author[5]{M.~Ardid}
\author[5]{S.~Ardid}
\author[6]{J.-J.~Aubert}
\author[7]{J.~Aublin}
\author[7]{B.~Baret}
\author[8]{S.~Basa}
\author[9]{B.~Belhorma}
% ...

\affiliation[1]{\scriptsize{Universit\'e de Strasbourg, CNRS,  IPHC UMR 7178, F-67000 Strasbourg, France}}
\affiliation[2]{\scriptsize Universit\'e de Haute Alsace, F-68100 Mulhouse, France}
\affiliation[3]{\scriptsize{IFIC - Instituto de F\'isica Corpuscular (CSIC - Universitat de Val\`encia) c/ Catedr\'atico Jos\'e Beltr\'an, 2 E-46980 Paterna, Valencia, Spain}}
\affiliation[4]{\scriptsize{Technical University of Catalonia, Laboratory of Applied Bioacoustics, Rambla Exposici\'o, 08800 Vilanova i la Geltr\'u, Barcelona, Spain}}
\affiliation[5]{\scriptsize{Institut d'Investigaci\'o per a la Gesti\'o Integrada de les Zones Costaneres (IGIC) - Universitat Polit\`ecnica de Val\`encia. C/  Paranimf 1, 46730 Gandia, Spain}}
\affiliation[6]{\scriptsize{Aix Marseille Univ, CNRS/IN2P3, CPPM, Marseille, France}}
\affiliation[7]{\scriptsize{Universit\'e Paris Cit\'e, CNRS, Astroparticule et Cosmologie, F-75013 Paris, France}}
\affiliation[8]{\scriptsize{Aix Marseille Univ, CNRS, CNES, LAM, Marseille, France }}
\affiliation[9]{\scriptsize{National Center for Energy Sciences and Nuclear Techniques, B.P.1382, R. P.10001 Rabat, Morocco}}
% ...

\emailAdd{dornic@cppm.in2p3.fr}

%% file: introduction.tex

Multi-messenger approaches consisting \color{blue}of \color{black}  concomitant searches for the same sources with neutrino telescopes, gravitational wave interferometers and/or multi-wavelength facilities constitute a privileged way of identifying astrophysical cosmic-ray accelerators. 
Neutrino astronomy allows the study of the most energetic non-thermal processes in the Universe and provides insight into source characteristics not accessible through other messengers. By constantly monitoring at least one complete hemisphere of the sky, neutrino telescopes are \color{blue}well-designed \color{black} to detect neutrinos emitted by transient phenomena. Real-time searches with the ANTARES telescope have been performed to look for neutrino candidates coincident with gamma-ray bursts detected by the Swift and Fermi satellites, high-energy neutrino events registered by IceCube, transient events from blazars monitored by HAWC, photon-neutrino coincidences by AMON notices and gravitational wave candidates observed by LIGO/Virgo. Requiring a spatial and temporal coincidence with other messengers increases the sensitivity and the significance of a potential discovery with respect to a solely neutrino based search.\\

The ANTARES telescope, completed in 2008, was the first operating neutrino telescope in the Mediterranean Sea~\cite{Collaboration:2011nsa}. It was composed of 12 detection lines of about 500 m height anchored at 2475 m depth offshore Toulon (42$^\circ$48'N, 6$^\circ$10'E). The mean distance between lines was about 65 m. Each line was made \color{blue} of \color{black} 25 storeys with an inter-storey distance of 14.5 m. Every storey \color{blue} held \color{black} three optical modules housing a single 10-inch diameter photomultiplier tube (PMT) looking downward at an angle of 45$^{\circ}$. In total, a $\sim$~10 Mt mass of water was instrumented with 885 optical modules. ANTARES had an average \color{blue} data-taking \color{black} efficiency larger than 94~\% with an effective area decreasing in time due to a loss of optical modules in operation. The data losses can be due to the shutdown of \color{blue} data-taking \color{black}, calibration periods or too high bioluminescence activities. The ANTARES data acquisition was switched off mid February 2022, when KM3NeT~\cite{KM3Net:2016zxf} reached a comparable instantaneous sensitivity to cosmic neutrinos.\\

\begin{figure}[!h]
    \centering
    \includegraphics[width=0.75\linewidth]{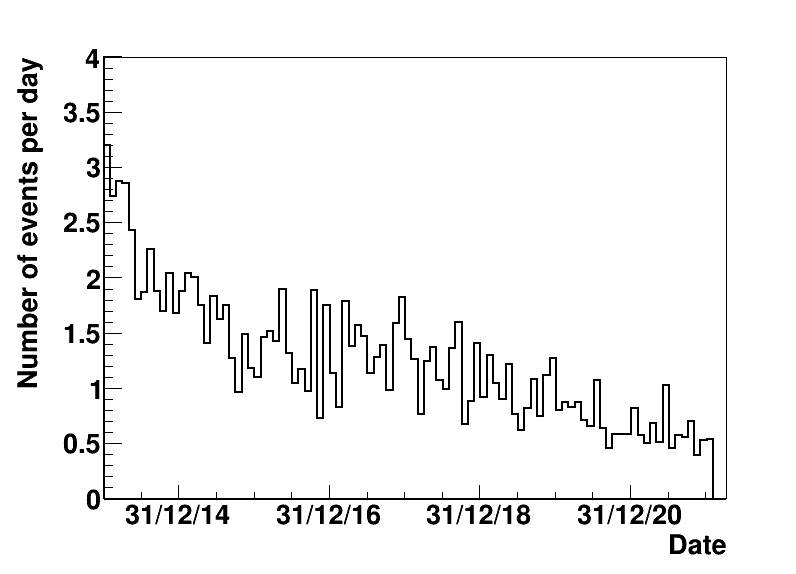}
    \caption{Evolution of the average number of neutrinos per day (averaged during a month period) between 2014 and 2022 selected for the ANTARES online analysis.}
    \label{fig:numu}
\end{figure}

In the online analysis framework, a dedicated real-time pipeline was developed to look for neutrino events in both \color{blue} temporal and spatial \color{black} coincidence with transient events announced by public alerts distributed through the Gamma-ray Coordinates Network (GCN~\url{https://gcn.gsfc.nasa.gov}) or by private alerts transmitted via special channels (i.e. special private requests from external communities). This analysis framework also hosted a neutrino alert sending program~\cite{2016JCAP...02..062A}.
The online selection was optimised to yield a neutrino sample (atmospheric and cosmic) with a minimal contamination from atmospheric muons. Similar cuts as the ones designed for the ANTARES standard offline point-like source search~\cite{2021ApJ...911...48A} were applied here: only upgoing track events with a good reconstruction quality are used in the analysis to ensure a median angular resolution of about 0.5$^{\circ}$. This leads to an atmospheric muon contamination lower than 10~\%~\cite{Ageron:2011pe}. The typical rate of neutrino candidates after selection in ANTARES is shown in Figure~\ref{fig:numu}, revealing a slow continuous efficiency loss of the detector~\cite{2018arXiv180508675A}. The online analysis used an ideal static detector both for the trigger and the reconstruction. It did not include \color{blue} knowledge \color{black} of the dynamical positioning and the precise charge and time calibration sets, which were made available a few months later for the offline analysis. \\

For interesting cases, more optimised offline analyses, using the most precise knowledge of the detector, are then performed to improve the online search~\cite{ANTARES:2017bia}. Offline searches are also performed for neutrino counterparts to catalogued transients~\cite{Molla:2020wsw,Adrian-Martinez:2016xij}. \\

 This paper focuses on the outcomes of the real-time follow-up program of ANTARES, in operation since 2014. Sections 2 and 3 summarise the results for the triggers provided by IceCube and LIGO/Virgo. Sections 4 and 5 present the follow-up of electromagnetic (EM) transients for gamma-ray bursts (GRB) and transient alerts reported by HAWC, respectively. Conclusions and outlooks are drawn in Section 6.

%% file: IC_follow-up.tex
A detection of neutrinos by ANTARES and IceCube telescopes within a close temporal window and with compatible directions (coincidence) would be \color{blue} strong evidence \color{black} of their astrophysical origin and would point directly to the position of the source in the sky. An alert for a neutrino coincidence would be so rare \color{blue} that the astronomy community would be motivated to perform a prompt \color{black} and multi-frequency EM follow-up.


Since 2016, IceCube has been sending public triggers~\cite{2021ApJ...910....4A} for high-energy starting events (HESE) and extremely high-energy track candidates (EHE). The events are received by the Astrophysical Multi-messenger Observatory Network (AMON~\cite{Smith:2012eu}) and distributed to the community via an alert of the GCN. In June 2019, IceCube substituted these alerts with two new \color{blue} very-high-energy \color{black} track event samples: gold (with a probability to be astrophysical $>$ 50~\%) and bronze ($>$ 30~\%) samples~\cite{2021ApJ...910....4A}. In July 2020, IceCube provided another alert stream based on very high-energy cascades with a typical resolution uncertainty of 15-20$^{\circ}$ (50 \% radius) and a typical rate of 8 events per year. The list of triggers is available at this address: \url{https://gcn.gsfc.nasa.gov/amon.html}.\\

\begin{figure}[!h]
\centering
  \includegraphics[width=0.75\linewidth]{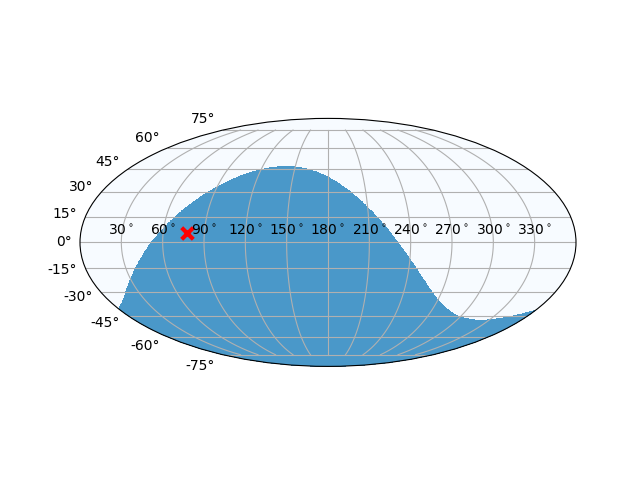}
  \caption{Sky map in equatorial coordinates showing in blue the visible ANTARES field of view at the time of IC170922. The direction of the IC event is drawn as a red cross.}
  \label{fig:IC170922}
\end{figure}

In this context, follow-up analyses have been performed for each IceCube event \color{blue} with a position on the sky below \color{black} the horizon of ANTARES (which could consequently yield an upgoing event at the time of the alert). ANTARES has received 115 neutrino triggers from the IceCube alert system and has followed 37 alerts (7 HESE, 3 EHE, 10 gold and 17 bronze). The rest of the triggers was either retracted by the IceCube Collaboration or located in the opposite hemisphere. As an illustration, Figure~\ref{fig:IC170922} shows the direction of the IceCube event IC170922 and the ANTARES visibility (i.e., the visible solid angle yielding upgoing events) at the time of the event. No neutrino candidates were found within a cone of 3$^{\circ}$ centred on the IceCube event coordinates and a time window of $\pm$1 hour, further extended to $\pm$1 day. These non-detections have been used to derive preliminary 90 \% confidence level (C.L.) upper limits on the radiant neutrino fluence\footnote{The radiant fluence, F, is computed with the formula: $F=\int_{E{\rm min}} ^{E_{\rm max}} E\phi_{0}  \left(\frac{E}{E_0}\right)^{-\gamma} dE$, where $\phi_{0}$ refers to the normalisation of the neutrino spectrum,  $\gamma$ is the spectral index and $E_0=1$\,GeV. $[E_{min}; E_{max}]$ corresponds to the 5--95\% energy range of the detectable neutrino flux.} of the possible sources \color{blue} producing \color{black} these events of the order of $\sim$~15 GeV cm$^{-2}$ and $\sim$~30 GeV cm$^{-2}$ for the assumed $E^{-2}$ and the $E^{-2.5}$ differential neutrino fluxes, respectively (see columns 3 and 4 in Table~\ref{tab:IC_limits}). These results have been published as GCN circulars and Astronomer's Telegrams typically one day after the alerts (columns 5 and 6 in Table~\ref{tab:IC_limits}). 

\begin{table}[!h]{
\begin{center}{
\begin{tabular}{|c|c|c|c|c|c|}
\hline
IceCube event & Elevation  &  \multicolumn{2}{c|}{Fluence U.L. (GeV cm$^{-2}$) at 90~\% C.L.} & GCN & ATels\\
             &  & $dN/dE\propto E^{-2}$& $dN/dE\propto E^{-2.5}$ & Id & Id\\
\hline
IC160731A (EHE/HESE) & -28$^{\circ}$   &  14 (2.8 TeV - 3.1 PeV)  &  27 (0.4 - 280 TeV) 	& /    &  9324 \\
IC160814A (HESE)         & $-$26$^{\circ}$   &  16 (2.9 TeV - 3.3 PeV)  &  43 (0.5 - 250 TeV) 	& 19885 &  9440 \\
IC161103A (HESE)         & $-$26$^{\circ}$   &  13 (3.8 TeV - 3.8 PeV)  &  22 (0.7 - 370 TeV) 	& 20134 &  9715 \\
IC170321A (EHE)          & $-$57$^{\circ}$   &  16 (2.5 TeV - 2.5 PeV)  &  26 (0.5 - 220 TeV) 	& 20926 & 10189 \\
IC170922A (EHE)          & $-$14$^{\circ}$   &  15 (3.3 TeV - 3.4 PeV)  &  34 (0.5 - 280 TeV) 	& 21923 & 10773 \\
IC171015A (HESE)         & $-$45$^{\circ}$   &  14 (2.7 TeV - 2.9 PeV)  &  27 (0.4 - 240 TeV) 	& 22019 & 10854 \\
IC180908A (EHE)          & $-$41$^{\circ}$   &  18 (2.4 TeV - 2.6 PeV)  &  36 (0.4 - 250 TeV) 	& 23218 & 12024 \\
IC190104A (HESE)         & $-$39$^{\circ}$   &  16 (3.2 TeV - 3.5 PeV)  &  30 (0.6 - 320 TeV) 	& 23611 & 12359 \\
IC190124A (HESE)         & $-$44$^{\circ}$   &  15 (3.1 TeV - 3.6 PeV)  &  25 (0.6 - 320 TeV) 	& 23793 & 12423 \\
IC190504A (HESE)         & $-$18$^{\circ}$   &  16 (3.1 TeV - 3.5 PeV)  &  32 (0.6 - 320 TeV) 	& 24400 & 12731 \\
IC190619A (gold)         & $-$19$^{\circ}$   &  13 (3.9 TeV - 3.9 PeV)  &  33 (0.7 - 320 TeV) 	& 24866 & 12878 \\
IC190712A (bronze)       & $-$13$^{\circ}$   &  16 (4.6 TeV - 4.3 PeV)  &  40 (0.8 - 420 TeV) 	& 25064 & 12937 \\
IC191119A (gold)         & $-$37$^{\circ}$   &  16 (3.4 TeV - 3.6 PeV)  &  28 (0.7 - 340 TeV) 	& 26266 & 13295 \\
IC191231A (bronze)       & $-$17$^{\circ}$   &  15 (5.3 TeV - 5.0 PeV)  &  32 (1.0 - 470 TeV) 	& 26623 & 13380 \\
IC200127A (bronze)       & $-$18$^{\circ}$   &  15 (6.8 TeV - 6.3 PeV)  &  29 (1.0 - 610 TeV) 	& 26811 & 13409 \\
IC200421A (bronze)       & $-$25$^{\circ}$   &  15 (3.9 TeV - 4.9 PeV)  &  27 (0.7 - 380 TeV) 	& 27619 & 13654 \\
IC200530A (gold)         & $-$0.04$^{\circ}$   &  80 (6.0 TeV - 6.0 PeV)  &  110 (1 - 560 TeV) 	& 27871 & 13770 \\
IC200620A (bronze)       & $-$32$^{\circ}$   &  15 (5.0 TeV - 4.0 PeV)  &  30 (0.8 - 400 TeV) 	& 28002 & 13820 \\
IC200911A (bronze)       & $-$7$^{\circ}$   &  14 (10.0 TeV - 8.0 PeV)  &  34 (1.5 - 740 TeV) 	& 28415 & 14008 \\
IC200916A (bronze)       & $-$29$^{\circ}$   &  18 (4.0 TeV - 4.5 PeV)  &  33 (1 - 430 TeV) 	& 28446 & 14025 \\
IC200926B (bronze)  & $-$13$^{\circ}$   &  15 (8.0 TeV - 7.0 PeV)  &  35 (1 - 690 TeV) 		& 28515 & 14045 \\
IC200929A (gold)    & $-$10$^{\circ}$   &  13 (3.0 TeV - 4.0 PeV)  &  45 (0.7 - 340 TeV) 		& 28535 & 14054 \\
IC201014A (bronze)  & $-$30$^{\circ}$   &  18 (5.0 TeV - 4.5 PeV)  &  30 (0.8 - 430 TeV) 		& 28624 & 14095 \\
IC201021A (bronze)  & $-$14$^{\circ}$   &  19 (5.0 TeV - 5.0 PeV)  &  48 (0.8 - 430 TeV) 		& 28738 & 14110 \\
IC201114A (bronze)  & $-$41$^{\circ}$   &  19 (4.0 TeV - 4.0 PeV)  &  31 (0.7 - 370 TeV) 		& 28890 & 14176 \\
IC201115A (gold)    & $-$7$^{\circ}$   &  17 (3.2 TeV - 3.2 PeV)  &  68 (0.6 - 330 TeV) 		& 28901 & 14181 \\
IC201209A (gold)    & $-$34$^{\circ}$   &  16 (3.0 TeV - 3.0 PeV)  &  30 (0.5 - 280 TeV) 		& 29023 & 14259 \\
IC210210A (gold)    & $-$18$^{\circ}$   &  16 (3.5 TeV - 3.7 PeV)  &  40 (0.7 - 360 TeV) 		& 29475 & / \\
IC210922A (gold)    & $-$37$^{\circ}$   &  16 (3.0 TeV - 3.3 PeV)  &  30 (0.6 - 300 TeV) 		& 30875 & 14935 \\
IC210926A (cascade) & $-$71$^{\circ}$   &  21 (2.3 TeV - 3.2 PeV)  &  30 (0.4 - 240 TeV) 		& 30887 & 14938 \\
IC211023A (bronze)  & $+$0.8$^{\circ}$   &  12 (3.0 TeV - 3.0 PeV)  &  48 (0.6 - 300 TeV) 		& 30971 & 14995 \\
IC211116A (bronze)  & $-$47$^{\circ}$   &  16 (3.0 TeV - 3.5 PeV)  &  26 (0.7 - 320 TeV) 		& 31090 & 15042 \\
IC211117A (gold)    & $-$12$^{\circ}$   &  15 (3.0 TeV - 3.5 PeV)  &  43 (0.6 - 320 TeV) 		& 31094 & 15044 \\
IC211125A (bronze)  & $-$6$^{\circ}$   &  12 (5.0 TeV - 5.0 PeV)  &  35 (1 - 500 TeV) 		& 31128 & 15065 \\
IC211208A (bronze)  & $-$10$^{\circ}$   &  17 (5.0 TeV - 5.0 PeV)  &  43 (1 - 500 TeV) 		& 31225 & 15106 \\
IC211216A (bronze)  & $-$8$^{\circ}$   &  16 (5.0 TeV - 5/0 PeV)  &  49 (1 - 450 TeV) 		& 31252 & 15121 \\
IC211216B (bronze)  & $-$4$^{\circ}$   &  17 (5.0 TeV - 5.0 PeV)  &  40 (1 - 450 TeV) 		& 31262 & 15127 \\
IC220205B (gold)    & $-$51$^{\circ}$   &  16 (3.0 TeV - 3.3 PeV)  &  30 (0.6 - 300 TeV) 		& 31556 & 15207 \\
 \hline
\end{tabular}}
\caption{Upper limits (at 90\% C.L.) on neutrino fluence as derived from the non-observation of ANTARES coincidences for each IceCube neutrino candidate. For each upper limit, the energy range in which 90\% of events are observed (excluding the 5\% of events with the lower/higher energies) is given. The publication reference number (Id) in GCN and in Astronomer's Telegram of the ANTARES follow-up of IceCube public triggers are given in the last two columns.}
\label{tab:IC_limits}
\end{center}}
\end{table}

Given the importance of some of the IceCube alerts, dedicated offline analyses have been performed for the following events: IC170922A and the blazar TXS0506+056~\cite{2018ApJ...863L..30A}, \color{blue} and \color{black} IC191001A / IC200530A and the tidal disruption events AT2019dsg / AT2019fdr~\cite{2021ApJ...920...50A}. An offline time and space correlation analysis for 54 IceCube high-energy track-like neutrino events was performed with the ANTARES neutrino offline data set, resulting in no significant coincidences~\cite{2019ApJ...879..108A}.\\

%% file: GW_follow-up.tex
Current modeling of the binary black-hole merger evolution does not imply EM or neutrino counterparts. However, in a sufficiently dense circumbinary region, an accretion disk might form and/or a relativistic jet connected to the accretion could be released. Accreting black holes can drive relativistic outflows~\cite{Meszaros:2006rc}. In this case, the process might lead to gamma-ray emission with a potential high-energy neutrino counterpart if a hadronic component is present ~\cite{2016PhRvD.93l3011M, Kotera:2016dmp, Perna:2016jqh, Bartos:2016dgn, 2017MNRAS.464..946S, 2016ApJ...822L...9M}. More GW detections will \color{blue} probe \color{black} poorly known systems, \color{blue} i.e. those \color{black} with asymmetric masses, or very large masses, thus leaving room for possible discoveries. An EM counterpart, presumably associated with \color{blue} hadronic \color{black}  emission is more likely from neutron star/black hole (NSBH) or neutron star/neutron star (BNS) mergers. Most of the models are based either on the formation of a gamma-ray burst~\cite{PhysRevD.98.043020, 2017ApJ848L4K, Kimura:2019ipr} or a magnetar~\cite{Fang:2017tla}. The other advantage provided by neutrino follow-up is that the angular resolution of ANTARES~\cite{2016JCAP...02..062A} ($\sim$~0.5$^\circ$ at $\sim$~10 TeV) compared to the size of the gravitational wave error box (a few hundreds of square degrees on the sky) offers the possibility to drastically reduce the size of the region of interest in case of a coincident neutrino detection. \\

\begin{figure}[!h]
\centering
  \includegraphics[width=0.75\linewidth]{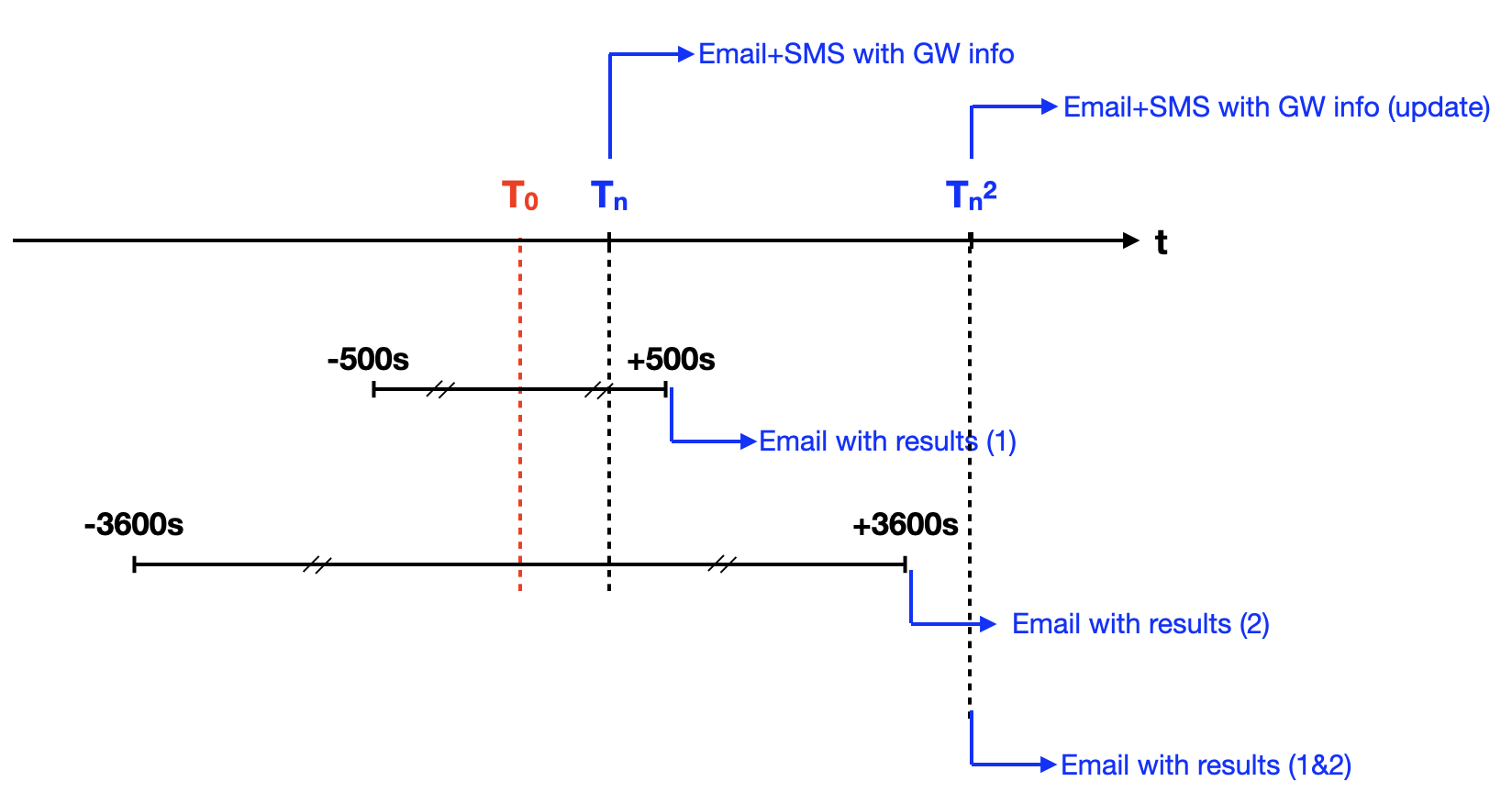}
  \caption{Principle of the online GW analysis. T$_{0}$, T$_{n}$ and T$_{n^2}$ correspond to the time of the detection, the time of the reception of the first notice and the time of the successive notices for one GW trigger. At the time T$_{n}$ and when the results of the two searches are available, one email is sent to the ANTARES GW subgroup. An SMS is also sent to speed up communication within the dedicated analysis group.}
  \label{fig:GWana}
\end{figure}

During the first observing run O1 in 2015, three GW events coming from binary black hole (BBH) mergers were detected by the LIGO interferometers~\cite{Abbott:2016blz}. As the GW online analysis was not ready at that time, only offline analyses have been performed by the ANTARES Collaboration~\cite{Adrian-Martinez:2016xgn, ANTARES:2017iky}. About  one year later, during the second observing run O2 (November 30, 2016 to August 25, 2017), the upgraded LIGO and Virgo detectors observed GWs from seven binary black hole mergers (plus 3 additional sources found in the offline analyses) and the BNS merger GW170817. Only for this last event, EM counterparts have been identified as a short gamma-ray burst followed by a kilonova~\cite{GBM:2017lvd, ANTARES:2017bia}. During the \color{blue} O2 run \color{black}, the LIGO/Virgo Collaboration triggered 15 alerts identified by \color{blue} online \color{black} analysis using a loose false-alarm-rate threshold of one per month. These triggers were shared with partner collaborations having signed a Memorandum of Understanding with LIGO/Virgo. Each of these alerts were followed by the ANTARES neutrino telescope by searching for a potential neutrino counterpart. The online analysis consists \color{blue} of \color{black} looking for (\textit{i}) temporal coincidences within $\pm$500~seconds and $\pm$1 hour time windows around the GW alert~\cite{Baret:2011tk} and (\textit{ii}) spatial overlap between the 90~\% probability contour from GW interferometers and the ANTARES visibility region at the time of the GW event. Figure~\ref{fig:GWana} illustrates the principle of the real-time GW analysis. LIGO and Virgo are sending a few notices for each GW candidate with updated information (Preliminary, Initial, Update, Retraction). Each new type is processed as a new GW trigger. At the end, the results of the last stable revision are provided. This analysis scheme has been applied to all the GW candidate triggers: no upgoing neutrino candidates temporally coincident with any of the GW candidates were found. The results of the nearly real-time analyses have been transmitted to \color{blue} the \color{black} LIGO/Virgo follow-up community via the GCN. Table~\ref{tab:GWresults0} lists the different GCN circulars sent on behalf \color{blue} of \color{black} the ANTARES Collaboration. In general, the online analyses performed for each GW candidate have been followed by a more optimised all-sky analysis~\cite{ANTARES:2017bia, Molla:2020wsw}. \\


\begin{table}[h!]{
\begin{center}{
\begin{tabular}{|c|c|c|c|c|c|}
\hline
  GW alert & Confirmed & Type &Fluence U.L. & Fluence U.L. & GCN Id\\
   & GW name &  & $E^{-2}$ & $E^{-2.5}$ & \\
 \hline
G268556  & GW170104 & BBH & 12 - 122 (3.6 TeV - 3.9 PeV)  &  20 - 756 (0.6 - 350 TeV) & 20517 \\ 
G270580  & & BBH & 13 - 48 (4.0 TeV - 4.0 PeV)  &  19 - 193 (0.7 - 370 TeV) & 20621 \\
G274296  & & BBH & 13 - 20 (2.8 TeV - 2.7 PeV)  &  25 - 79 (0.5 - 270 TeV) & 20704 \\ 
G275404  & & BBH & 14 - 49 (4.1 TeV - 4.4 PeV)  &  19 - 174 (0.7 - 390 TeV) & 20751 \\
G275697  & & BBH & 12 - 25 (3.3 TeV - 3.7 PeV)  &  22 - 60 (0.5 - 330 TeV) & 20765 \\ 
G277583  & & BBH & 13 - 100 (3.3 TeV - 3.5 PeV)  &  22 - 477 (0.6 - 310 TeV) & 20866 \\
G284239  & & BBH & 13 - 84 (3.2 TeV - 3.4 PeV)  &  22 - 448 (0.5 - 310 TeV) & 21066 \\ 
G288732  & GW170608 & BBH & 14 - 17 (5.4 TeV - 5.4 PeV)  &  20 - 53 (0.9 - 490 TeV) & 21223 \\
G296853  & GW170809 & BBH & 14 - 17 (2.4 TeV - 2.8 PeV)  &  35 - 66 (0.4 - 240 TeV) & 21433 \\ 
G297595  & GW170814 & BBH & 13 - 16 (2.4 TeV - 2.8 PeV)  &  33 - 64 (0.4 - 240 TeV) & 21479 \\
G298048  & GW170817 & BNS & & & 21522 / 21631 \\ 
G298936  & GW170823 & BBH & 12 - 21 (3.0 TeV - 3.5 PeV)  &  21 - 68 (0.5 - 300 TeV) & 21659 \\
G299232  & & BBH & 15 - 24 (4.3 TeV - 4.5 PeV)  &  19 - 85 (0.7 - 400 TeV) & 21696 / 21769 \\
   \hline
\end{tabular}}
\caption{ANTARES analysis results of the GW candidates distributed by LIGO/Virgo during O2 run. Columns 3 and 4 provide the upper limits (at 90~\% C.L.) on neutrino fluence as derived from the non-observation of ANTARES coincidences in the upgoing sky for the GW candidates distributed by LIGO/Virgo during run O3 assuming a neutrino energy spectrum of E$^{-2}$ and E$^{-2.5}$. The range in the upper limits corresponds to the minimum and maximum values, depending on the local coordinates. Note that for the event G298048, all the GW provenience area is in the downgoing sky. For each upper limit, the energy range in which 90~\% of events are observed (excluding the 5~\% of events with the lower/higher energies) is given. The last column gives the references of the GCN circular published by the ANTARES collaboration for each GW candidates.}
\label{tab:GWresults0}
\end{center}}
\end{table}

\begin{figure}[!ht]
\centering
  \includegraphics[width=0.75\linewidth]{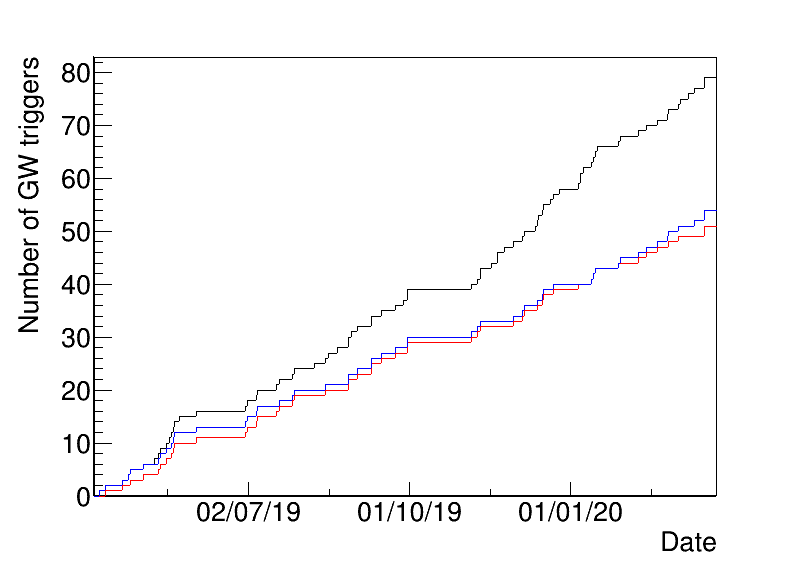}
  \caption{Cumulative number of the GW candidates detected during \color{blue} the O3 run \color{black} as a function of the date: all GW triggers (black); analysable triggers, i.e., not terrestrial, nor retracted at the time of the analysis (blue); analysed GW candidates (red) by ANTARES.}
  \label{fig:GWclass}
\end{figure}

\begin{figure}[!h]
\centering
  \includegraphics[width=0.75\linewidth]{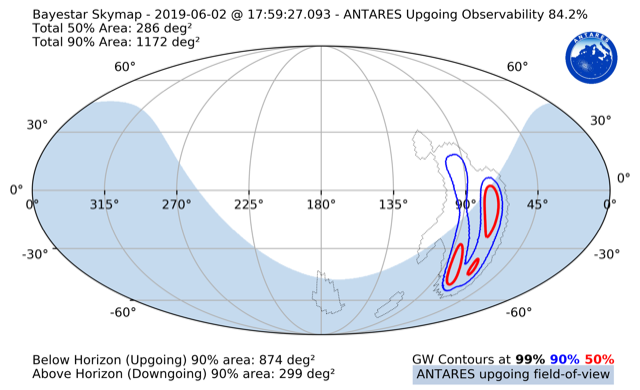}
  \caption{Sky map in equatorial coordinates showing the 99 \% (gray), 90 \% (blue) and 50 \% (red) probability contours for the allowed provenience region of S190602aq together with the ANTARES field of view at the event time (blue part of the
map).}
  \label{fig:GW190602}
\end{figure}

The third observing run O3 started on April 1st, 2019, with \color
{blue} even-more-upgraded \color{black} interferometers. Until the end of March 2020, 78 alerts were distributed publicly, \color{blue} with \color{black} 22 retracted by LIGO/Virgo after further investigations. Among the 56 events, 37 are classified as BBH, 5 Mass Gap, 4 NSBH, 6 BNS, 1 unmodeled and 3 probably coming from terrestrial noise. The real-time analysis has been performed for 51 GW triggers (Figure~\ref{fig:GWclass}). Two triggers have not been analysed since the GW were quickly classified as terrestrial noise, 2 had their allowed provenience regions totally outside the ANTARES field of view while for the other events, ANTARES was in maintenance. As an example, Figure~\ref{fig:GW190602} illustrates the probability contours of the GW event S190602aq together with the ANTARES visibility at the time of the event. The main characteristics of the 51 GW candidates and the results of the neutrino search are summarised in Table~\ref{tab:GWresults}. 

\begin{figure}[!h]
\centering
  \includegraphics[width=0.75\linewidth]{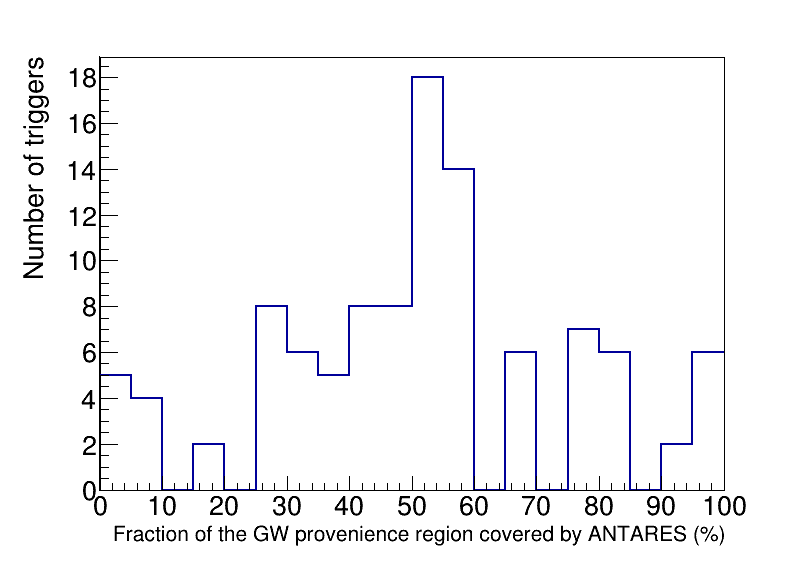}
  \caption{Distribution of the fraction of the 90~\% C.L. allowed provenience region of the GW candidates visible by ANTARES at T$_{0}$ as upgoing directions for neutrino candidates.. }
  \label{fig:GWfrac}
\end{figure}

Figure~\ref{fig:GWfrac} shows the distribution of the fraction of the 90 \% C.L. allowed provenience region visible by the ANTARES detector at the GW detection time, T$_{0}$, as upgoing directions for all the GW triggers. Unfortunately, during O3, most of the GW candidates were reconstructed with a large allowed provenience region, typically above 1000 deg$^{2}$. Only a few events were reconstructed with a provenience region of less than 100 deg$^{2}$. This makes the EM follow-up even more difficult and the detection of a neutrino more relevant.\\

\begin{figure}[!h]
\centering
  \includegraphics[width=0.65\linewidth]{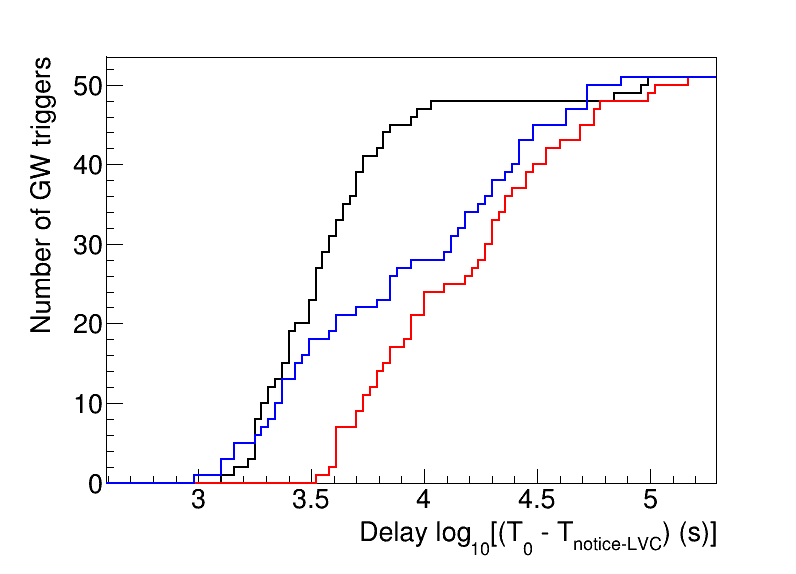}
  \caption{Cumulative distribution of time delays between the time of the 51 GW candidates and the reception time of the first notice (black) and the submission times of the circular with the ANTARES results (red). The blue curve corresponds to the time difference between the reception of the GW confirmation circular and the ANTARES circulars.}
  \label{fig:GWtime}
\end{figure}

For those events detected during the O3 period of GW interferometers, similar analyses to the ones for the O2 events have been performed, but in a completely automated way. No tracks induced by a muon neutrino have been found in time and space coincidence with any alert of run O3 (Table~\ref{tab:GWresults}). \color{blue} From \color{black} the non-observation of ANTARES coincidences, upper limits on the neutrino fluence have been estimated (Table~\ref{tab:GWlimits}). All the results have been reported via the publication of a circular to the GCN. The provided information contains a sky map of the visible region of ANTARES (as upgoing) at the time of the GW candidate together with the GW allowed provenience regions (see Figure~\ref{fig:GW190602} as an example), the fraction of the GW 90 \% allowed provenience region covered by the ANTARES field of view, the number of detected events in time/space coincidence and the expected number of atmospheric background events in the region visible by ANTARES. This expected background rate is computed directly from the data using an off-region area before the GW trigger. The results are reported for two search time windows: $\pm$500 s and $\pm$1 hour centred on the time of the GW alert. The typical delay between the detection time of each GW candidate and the time of the ANTARES circular is \color{blue} about \color{black} 4.5 hours (Figure~\ref{fig:GWtime}). Note that one necessary condition to submit our results was to receive a confirmation circular by LIGO/Virgo. If this time is used as a reference, the results have been published on average in less than 2~h (Figure~\ref{fig:GWtime}). The offline analysis has already been done for a few selected events published by LIGO/Virgo~\cite{Abbott:2020uma, LIGOScientific:2020stg, ColomerMolla:2020dln}.\\

\label{tab:GWresults}
\begin{longtable}{|c|c|c|c|c|c|}
\caption{Characteristics of the GW candidates distributed by LIGO/Virgo during run O3. The coverage indicates the fraction of the 90~\% GW allowed provenience region falling in the visibility of ANTARES at the time of the event. The allowed area is the size of the region provided by GW interferometers at the time of the analysis. Some more refined GW parameters may have arrived later. The GW are ordered by date (YYMMDD), contained in the name of the GW (as in GraceDB \url{https://gracedb.ligo.org/superevents/public/O3/}). The GCN references are the ones published by the ANTARES Collaboration.}
\\
\hline 
GW name & Type & Allowed area & Distance & Coverage & GCN\\ 
  &  & (deg$^{2}$) & (Mpc) &  (\%) & Id\\ 
\hline
\endfirsthead

 \hline 
\endfoot

\multicolumn{6}{c}{{\bfseries \tablename\ \thetable{} -- (continued)}} \\
\hline
GW name & Type & Allowed area & Distance & Coverage & GCN\\ 
\hline
\endhead

\hline 
\endlastfoot

S190412m  & BBH &  156 &  812  & 9.3   &   24105   \\ 
S190421ar  & BBH &  1917 &  1628  & 52   &   24156   \\ 
S190426c  & BNS &  1932 &  377  & 45   &   24271   \\ 
S190503bf  & BBH &  448 &  421  & 98   &   24387   \\ 
S190512at  & BBH &  399 &  1388  & 83   &   24516   \\ 
S190513bm  & BBH &  691 &  1987  & 55   &   24539   \\ 
S190517h  & BBH &  939 &  2950  & 83.3   &   24581   \\ 
S190519bj  & BBH &  967 &  3154  & 34   &   24602   \\ 
S190521g  & BBH &  1163 &  3931  & 56   &   24628   \\ 
S190521r  & BBH &  388 &  1136  & 30   &   24634   \\
S190602aq  & BBH &  1172 &  797  & 84   &   24719   \\ 
S190630ag  & BBH &  8493 &  926  & 68.6   &   24924   \\ 
S190701ah  & BBH &  67 &  1849  & 99.9   &   24952   \\ 
S190706ai  & BBH &  1100 &  5263  & 48.7   &   25009   \\ 
S190707q  & BBH &  1375 &  874  & 58.3   &   25013   \\ 
S190718y  & Terrestrial &  7246 &  227  & 77.5   &   25091   \\ 
S190720a  & BBH &  1599 &  869  & 41.6   &   25120   \\ 
S190727h  & BBH &  1357 &  2839  & 55.1   &   25168   \\ 
S190728q  & BBH &  977 &  874  & 38.1   &   25194   \\ 
S190814bv  & NSBH &  772 &  267  & 99.9   &   25330   \\ 
S190828j  & BBH &  603 &  1946  & 53.1   &   25508   \\ 
S190828l  & BBH &  948 &  1528  & 56.8   &   25507   \\ 
S190901ap  & BNS &  13613 &  241  & 54.1   &   25611   \\ 
S190910d  & NSBH &  3829 &  632  & 50.9   &   25700   \\ 
S190910h  & BNS &  24226 &  230  & 50   &   25711   \\ 
S190915ak  & BBH &  528 &  1584  & 45.6   &   25758   \\ 
S190923y  & MassGap &  2107 &  438  & 41.7   &   25816   \\ 
S190924h & MassGap & 515 & 514 & 39.6 & 25836 \\
S190930s  & MassGap &  1998 &  709  & 25.9   &   25881   \\ 
S190930t  & NSBH &  24220 &  108  & 50   &   25882   \\ 
S191105e  & BBH &  1253 &  1183  & 66.4   &   26189   \\ 
S191109d  & BBH &  1487 &  1810  & 79.1   &   26210   \\ 
S191110af  & unmodeled &  1261 &    & 56   &   26230   \\ 
S191129u  & BBH &  1011 &  742  & 51.6   &   26307   \\ 
S191204r  & BBH &  433 &  678  & 92.9   &   26336   \\ 
S191205ah  & NSBH &  6378 &  385  & 28   &   26352   \\ 
S191213g  & BNS &  1393 &  201  & 75.1   &   26404   \\ 
S191215w  & BBH &  923 &  1770  & 42.3   &   26443   \\ 
S191216ap  & BBH &  300 &  376  & 15.9   &   26458   \\ 
S191222n  & BBH &  2324 &  2518  & 53.1   &   26550   \\ 
S200105ae  & Terrestrial &  7719 &  283  & 35.6   &   26643   \\ 
S200112r  & BBH &  6199 &  1125  & 38.4   &   26718   \\ 
S200114f  & Unmodeled &  403 &    & 6   &   26742   \\ 
S200115j  & MassGap &  920 &  340  & 76   &   26762   \\ 
S200128d  & BBH &  2521 &  3702  & 50.1   &   26912   \\ 
S200208q  & BBH &  1120 &  2142  & 68.5   &   27016   \\ 
S200213t  & BNS &  2587 &  201  & 31.9   &   27049   \\ 
S200219ac  & BBH &  1251 &  3533  & 55.5   &   27135   \\ 
S200225q  & BBH &  403 &  995  & 42.4   &   27201   \\ 
S200302c  & BBH &  6704 &  1820  & 49.5   &   27284   \\ 
S200316bj  & MassGap &  1117 &  1178  & 26.6   &   27390 \\ \hline   

\end{longtable}

\label{tab:GWlimits}
\begin{longtable}{|c|c|c|}
\caption{Upper limits (at 90~\% C.L.) on neutrino fluence as derived from the non-observation of ANTARES coincidences for the GW candidates distributed by LIGO/Virgo during run O3. The range in the upper limits corresponds to the minimum and maximum values, depending on the local coordinates. For each upper limit, the energy range in which 90~\% of events are observed (excluding the 5~\% of events with the lower/higher energies) is given. Some more refined GW parameters may have arrived later. The GW are ordered by date (YYMMDD), contained in the name of the GW (as in GraceDB \url{https://gracedb.ligo.org/superevents/public/O3/}). }
\\
\hline 
GW name & \multicolumn{2}{c|}{Fluence U.L. (GeV cm$^{-2}$) at 90~\% C.L.} \\
             &  $dN/dE\propto E^{-2}$& $dN/dE\propto E^{-2.5}$ \\
\hline
\endfirsthead

 \hline 
\endfoot

\multicolumn{3}{c}{{\bfseries \tablename\ \thetable{} -- (continued)}} \\
\hline
GW name & Fluence U.L. $E^{-2}$ & Fluence U.L. $E^{-2.5}$ \\ 
\hline
\endhead

\hline 
\endlastfoot

S190412m  & 14 - 16 (3.8 TeV - 3.8 PeV)  &  38 - 72 (0.6 - 350 TeV)   \\ 
S190421ar  & 13 - 21 (2.4 TeV - 1.9 PeV)  &  28 - 92 (0.5 - 220 TeV)   \\ 
S190426c  & 13 - 24 (6.1 TeV - 6.3 PeV)  &  18 - 60 (1.0 - 560 TeV)   \\ 
S190503bf  & 11 - 33 (2.4 TeV - 2.8 PeV)  &  36 - 145 (0.4 - 240 TeV)   \\ 
S190512at  & 15 - 54 (2.5 TeV - 2.9 PeV)  &  27 - 42 (0.4 - 250 TeV)   \\ 
S190513bm  & 15 - 51 (5.1 TeV - 5.4 PeV)  &  26 - 134 (0.9 - 480 TeV)   \\ 
S190517h  & 14 - 111 (2.5 TeV - 2.5 PeV)  &  32 - 718 (0.4 - 240 TeV)   \\ 
S190519bj  & 13 - 185 (4.1 TeV - 4.4 PeV)  &  22 - 785 (0.7 - 390 TeV)   \\ 
S190521g  & 14 - 111 (2.4 TeV - 2.5 PeV)  &  32 - 943 (0.4 - 230 TeV)   \\ 
S190521r  & 14 - 18 (3.0 TeV - 3.1 PeV)  &  31 - 64 (0.5 - 280 TeV)   \\
S190602aq  & 13 - 42 (2.6 TeV - 2.9 PeV)  &  31 - 190 (0.4 - 250 TeV)   \\ 
S190630ag  & 13 - 21 (2.7 TeV - 2.9 PeV)  &  25 - 83 (0.5 - 260 TeV)   \\ 
S190701ah  & 15 - 16 (3.1 TeV - 3.2 PeV)  &  23 - 28 (0.5 - 300 TeV)   \\ 
S190706ai  & 12 - 17 (3.7 TeV - 4.0 PeV)  &  27 - 56 (0.6 - 360 TeV)   \\ 
S190707q  & 13 - 34 (3.0 TeV - 3.3 PeV)  &  20 - 165 (0.5 - 290 TeV)   \\ 
S190718y  & 12 - 38 (4.3 TeV - 4.4 PeV)  &  17 - 2632 (0.7 - 410 TeV)   \\ 
S190720a  & 14 - 17 (2.4 TeV - 2.4 PeV)  &  28 - 83 (0.4 - 230 TeV)   \\ 
S190727h  & 14 - 16 (2.4 TeV - 2.1 PeV)  &  26 - 49 (0.5 - 230 TeV)   \\ 
S190728q  & 23 - 24 (4.1 TeV - 4.1 PeV)  &  86 - 87 (0.6 - 380 TeV)   \\ 
S190814bv  & 14 - 20 (2.4 TeV - 2.9 PeV)  &  51 - 69 (0.4 - 240 TeV)   \\ 
S190828j  & 12 - 39 (2.4 TeV - 2.8 PeV)  &  31 - 190 (0.4 - 240 TeV)   \\ 
S190828l  & 14 - 83 (2.4 TeV - 2.7 PeV)  &  37 - 733 (0.4 - 240 TeV)   \\ 
S190901ap  & 12 - 244 (3.3 TeV - 3.4 PeV)  &  22 - 200 (0.5 - 310 TeV)   \\ 
S190910d  & 13 - 54 (2.8 TeV - 2.6 PeV)  &  20 - 392 (0.5 - 260 TeV)   \\ 
S190910h  & 13 - 111 (3.0 TeV - 3.1 PeV)  &  20 - 562 (0.5 - 290 TeV)   \\ 
S190915ak  & 16 - 227 (8.6 TeV - 8.9 PeV)  &  23 - 366 (0.2 - 750 TeV)   \\ 
S190923y  & 13 - 131 (2.9 TeV - 2.7 PeV)  &  21 - 539 (0.5 - 270 TeV)   \\ 
S190924h  & 13 - 16 (3.2 TeV - 3.2 PeV)  &  46 - 75 (0.5 - 300 TeV)   \\ 
S190930s  & 13 - 40 (3.7 TeV - 3.6 PeV)  &  30 - 172 (0.6 - 340 TeV)  \\ 
S190930t  & 13 - 20 (3.2 TeV - 3.3 PeV)  &  21 - 936 (0.5 - 300 TeV)   \\ 
S191105e  & 14 - 18 (2.8 TeV - 3.2 PeV)  &  24 - 67 (0.5 - 270 TeV)   \\ 
S191109d  & 14 - 27 (2.6 TeV - 2.9 PeV)  &  22 - 72 (0.4 - 250 TeV)   \\ 
S191110af  & 12 - 110 (2.4 TeV - 2.6 PeV)  &  26 - 94 (0.4 - 240 TeV)  \\ 
S191129u  & 13 - 20 (2.7 TeV - 2.4 PeV)  &  22 - 80 (0.5 - 250 TeV)   \\ 
S191204r  & 15 - 16 (2.4 TeV - 2.9 PeV)  &  26 - 71 (0.4 - 240 TeV)   \\ 
S191205ah  & 12 - 297 (3.1 TeV - 3.3 PeV)  &  24 - 266 (0.5 - 300 TeV)   \\ 
S191213g  & 13 - 103 (2.9 TeV - 3.1 PeV)  &  23 - 733 (0.5 - 280 TeV)   \\ 
S191215w  & 14 - 188 (3.3 TeV - 3.3 PeV)  &  27 - 646 (0.5 - 310 TeV)   \\ 
S191216ap  & 13 - 30 (3.1 TeV - 3.2 PeV)  &  41 - 153 (0.5 - 290 TeV)   \\ 
S191222n  & 14 - 42 (2.8 TeV - 2.7 PeV)  &  22 - 226 (0.5 - 270 TeV)   \\ 
S200105ae  & 12 - 100 (3.5 TeV - 3.8 PeV)  &  22 - 764 (0.6 - 340 TeV)   \\ 
S200112r  & 12 - 134 (3.8 TeV - 4.0 PeV)  &  28 - 770 (0.6 - 360 TeV)   \\ 
S200114f  & 12 - 17 (2.4 TeV - 2.9 PeV)  &  37 - 46 (0.4 - 240 TeV)   \\ 
S200115j  & 13 - 136 (3.2 TeV - 3.5 PeV)  &  24 - 63 (0.5 - 310 TeV)  \\ 
S200128d  & 13 - 26 (3.0 TeV - 5.0 PeV)  &  22 - 127 (0.5 - 290 TeV)   \\ 
S200208q  & 19 - 21 (2.4 TeV - 2.9 PeV)  &  25 - 28 (0.4 - 240 TeV)   \\ 
S200213t  & 14 - 121 (5.3 TeV - 5.4 PeV)  &  15 - 290 (2.0 - 480 TeV)   \\ 
S200219ac  & 13 - 82 (3.3 TeV - 3.7 PeV)  &  24 - 310 (0.6 - 320 TeV)   \\ 
S200225q  & 12 - 23 (6.2 TeV - 6.2 PeV)  &  23 - 62 (1.1 - 570 TeV)  \\ 
S200302c  & 12 - 106 (3.5 TeV - 3.8 PeV)  &  23 - 380 (0.6 - 330 TeV)  \\ 
S200316bj  & 13 - 17 (2.4 TeV - 1.9 PeV)  &  25 - 64 (0.5 - 220 TeV) \\ \hline   
\end{longtable}

%% file: GRB_follow-up.tex
\begin{figure}[!h]
\centering
  \includegraphics[width=0.45\linewidth]{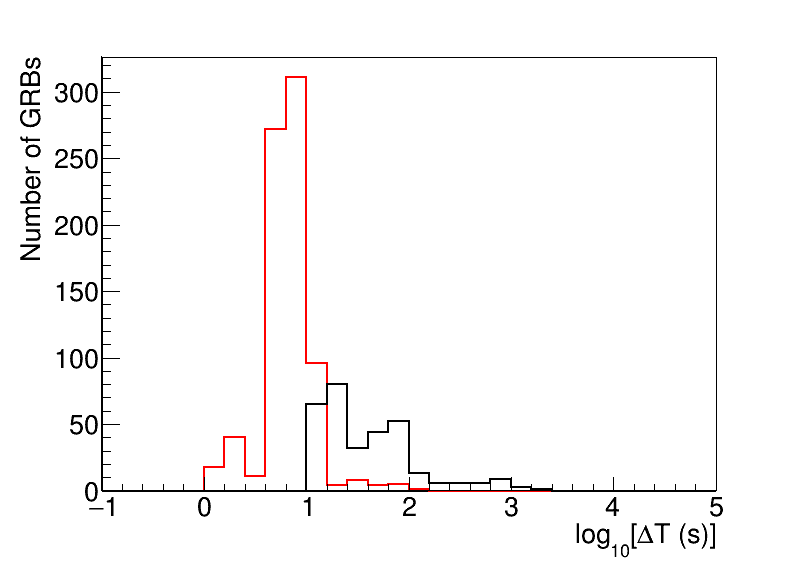}
  \includegraphics[width=0.45\linewidth]{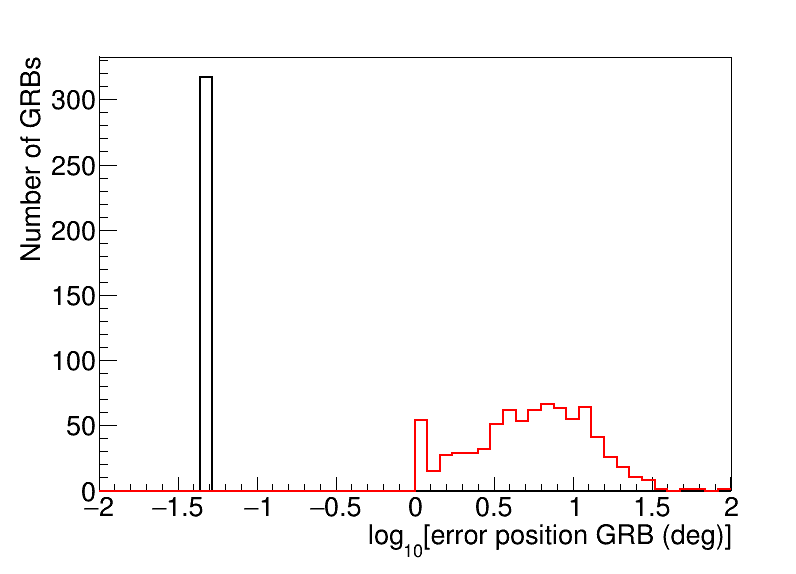}
  \caption{GRBs detected by Swift (black histogram) and Fermi (red histogram) selected in the ANTARES analysis from 01/2014 to 02/2022. (Left) Time delays, $\Delta T$, between the burst detection by Swift and Fermi and the received notice. (Right) Error in the position of the GRBs detected by Swift and Fermi.  The information \color{blue}is \color{black} extracted directly from the GCN notices.}
  \label{fig:GRBres}
\end{figure}
\begin{figure}[!h]
    \centering
    \includegraphics[width=0.75\linewidth]{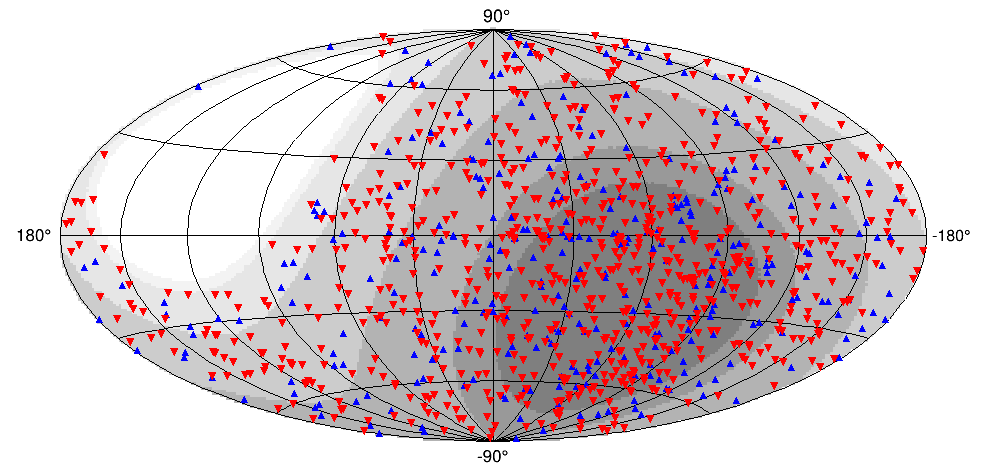}  
    \caption{Sky map in Galactic coordinates with the positions of the Fermi (red triangles) and Swift (blue triangles) GRBs followed by ANTARES in the full analysed period (01/2014 to 02/2022). The shade of grey indicates the ANTARES visibility. The darkest region indicates the maximum visibility.}
    \label{fig:GRBskymaps}
\end{figure}

Gamma-ray bursts are mainly detected by X-ray and gamma-ray satellites such as \color{blue} \textit{Swift} and \textit{Fermi} \color{black}. Once a GRB is detected, an alert message is sent publicly via the GCN within a few tens of seconds. Figure~\ref{fig:GRBres} left shows the delay of the alert sending for all GRBs detected by Swift and Fermi selected in the ANTARES analysis from 01/2014 to 02/2022 (see below for the details). ANTARES is able to react in real time to this type of alert. Only the bursts with directions below the ANTARES horizon are analysed online. A dedicated search for neutrino-induced muons in the online \color{blue} data-set \color{black} is performed in \color{blue} real-time \color{black} within a time window [--250~s; +750~s] around the detection time and in a cone centred on the GRB position. The radius of the cone is determined by taking the maximum between 2$^\circ$ (containing about 90~\% of the point spread function, see Figure~3 in Ref.~\cite{2017PhRvD..96h2001A}) and the size of the error box provided by Fermi (Figure~\ref{fig:GRBres} right). In the case of Swift triggers, a 2$^\circ$ cone is always used. For a cone radius of 2$^\circ$, the detection of one event yields a p-value (i.e., a probability that the coincidence is due to background) in the range of 2--5$\times$10$^{-5}$. The analysis is performed automatically. To ensure the quality of the data at the alert time, the detector stability is monitored over several hours before the alert, i.e., the reconstructed event rates should follow a Gaussian distribution. This analysis has been operational since the beginning of 2014 and $\sim$~98~\% of the alerts have been processed. Over more than 8 years of operation (01/2014--02/2022), there were 317 Swift and 770 Fermi-GBM bursts. The bursts detected at the same time by both satellites are tagged with the information provided by Swift. Figure~\ref{fig:GRBskymaps} shows the directions of the GRBs of both samples. No online neutrino signals have been detected in this search.

In \color{blue}the \color{black} case of a coincident neutrino detection (never found in our data), a dedicated offline analysis would have been used to confirm the result and to compute its significance (expected to be higher than 3$\sigma$ in most of the cases). Using the most precise knowledge of the detector, offline individual and stacked analyses on GRB catalogues have been performed with improved event selections~\cite{Adrian-Martinez:2016xij,Albert:2016eyr, 2021MNRAS.500.5614A}.\\

%% file: transient_follow-up.tex
\begin{figure}[!h]
    \centering
    \includegraphics[width=0.75\linewidth]{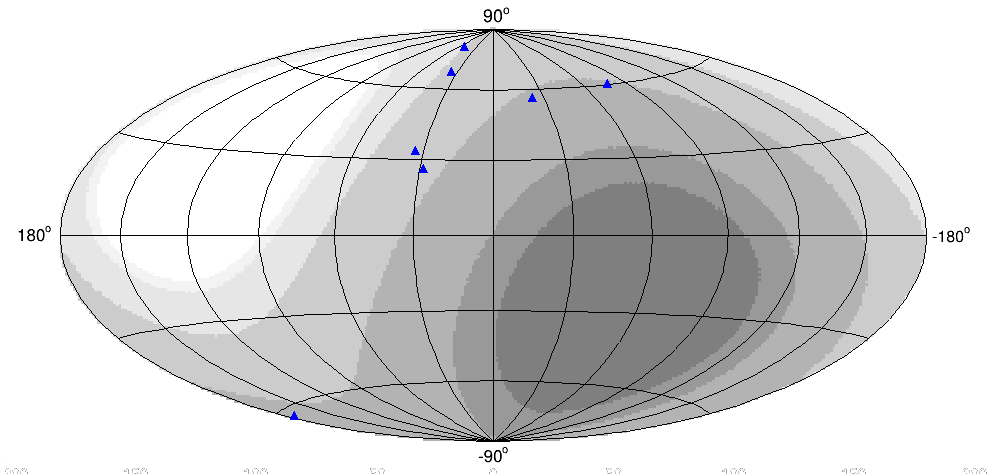}  
    \caption{Sky map in Galactic coordinates with the positions of the analysed HAWC alerts (blue triangles) up to Feb 2022. The shade of grey indicates the ANTARES visibility. The darkest region indicates the maximum visibility.}
    \label{fig:HAWCskymaps}
\end{figure}

Since mid 2019, the HAWC Collaboration has been issuing alerts of short TeV transients lasting from 0.2~s to 100~s, targeting in particular GRBs. HAWC shares the same advantage as \color{blue}ANTARES, being able to monitor half the sky with a high duty cycle \color{black}. The quest \color{blue}for TeV gamma rays produced \color{black} by transient astrophysical sources is particularly interesting for \color{blue}high-energy \color{black} neutrino telescopes. First, gamma ray detection proves that the sources generating the events are powerful cosmic accelerators. Second, in hadronic production scenarios, these gamma rays have almost the same flux and energy spectrum as the accompanying neutrinos, within the energy range in which the telescope is most sensitive.
The alerts are channeled via the AMON framework and then distributed by the GCN. Up to Feb. 2022, the HAWC Collaboration sent 22 triggers, 7 of them with a direction within the ANTARES field of view at the time of the alert. The alert parameters are available at this address: \url{https://gcn.gsfc.nasa.gov/amon_hawc_events.html}. Figure~\ref{fig:HAWCskymaps} shows the direction of the analysed HAWC alerts together with the integrated ANTARES visibility. The same analysis strategy as for the IceCube neutrino alerts is applied and the results are then published as a circular to the GCN and/or Astronomer's Telegram. No online neutrinos have been identified in coincidence with the HAWC transients. Table~\ref{tab:HAWC_publis} and Table~\ref{tab:HAWC_limits} summarise the main alert parameters, the GCN published and the corresponding upper limits for this analysis.

\begin{table}[h!]{
 \begin{center}{
\begin{tabular}{|c|c|c|c|c|c|}
\hline
Event Id		     &	Direction	   &	Error	   &	$\delta$T   & Visibility &  GCN        \\
			         &	RA, Dec (deg)	   &	(arcmin)   &	(s)	  & (\%)	 & 	 Id      \\
\hline
HAWC-211123A (110400\_42)	  & 34.12, $-$8.04     &   36         &    1       &   53  &   31111   \\ 
HAWC-210507A (1010067\_1345)	  & 257.24, $+$8.09    &	24		   &    10      &   45  &    29967  \\ 
HAWC-201019A (1009678\_72)	  & 203.15, $+$29.70   &   36         &    100     &   32  &    28729   \\ 
HAWC-200709A (1009500\_793)	  & 252.38, $+$15.20   &	24		   &    100	    &	42	&	 28074    \\
HAWC-200226A (19170\_50)	  &	182.80, $-$0.61    &	48     	   &	1	    &   50  &   27242     \\
HAWC-191210A (9066\_1171)     &	210.80, $-$1.52    &	48     	   &	0.2	    &   35  &   26391     \\
HAWC-191019A (8991\_1097)     & 217.50, $+$25.81   &	48     	   &	0.2	    &   50  &   26049     \\
\hline
\end{tabular}}
\caption{Summary of HAWC transient triggers and ANTARES publication in GCN. "Error" refers to the location uncertainty (radius, 50~\% containment). $\delta$T is the trigger duration interval. "Visibility" refers to the ANTARES visibility fraction of the alert direction during one day. The last column gives the GCN circular references of the ANTARES follow-up.}
\label{tab:HAWC_publis}
\end{center}}
\end{table}

\begin{table}{
\begin{center}{
\begin{tabular}{|c|c|c|c|c|c|}
\hline
Alert & \multicolumn{2}{c|}{Fluence U.L. (GeV cm$^{-2}$) at 90~\% C.L.} \\
             &  $dN/dE\propto E^{-2}$& $dN/dE\propto E^{-2.5}$ \\
\hline
HAWC-211123A & 17 (3 TeV - 3 PeV)  &  70 (0.5 - 280 TeV)  \\
HAWC-210507A & 15 (4 TeV - 4 PeV)  &  37 (0.7 - 380 TeV)  \\
HAWC-201019A & 17 (7 TeV - 6 PeV)  &  37 (1 - 630 TeV)  \\
HAWC-200709A & 40 (5 TeV - 5 PeV)  &  240 (0.9 - 430 TeV)  \\
HAWC-200226A & 16 (3.2 TeV - 3.4 PeV)  &  27 (0.6 - 320 TeV)  \\
HAWC-191210A & 17 (3 TeV - 3 PeV)  &  69 (0.6 - 300 TeV)  \\
HAWC-191019A & 90 (6 TeV - 6 PeV)  &  110 (1 - 550 TeV)  \\
\hline
\end{tabular}}
\caption{Upper limits (at 90~\% C.L.) on neutrino fluence as derived from the non-observation of ANTARES coincidences for each HAWC transient triggers. For each upper limit, the energy range in which 90~\% of events are observed (excluding the 5~\% of events with the lower/higher energies) is given.}
\label{tab:HAWC_limits}
\end{center}}
\end{table}

AMON \color{blue}also issues alerts for significant coincidences \color{black}between ANTARES neutrinos and Fermi/LAT photons and between IceCube neutrinos and HAWC transients. The ANTARES Collaboration has performed \color{blue}a \color{black} follow-up of the IceCube + HAWC coincidences (NuEM) with a similar analysis method as for the IceCube alerts. The NuEM alert parameters are available at this address: \url{https://gcn.gsfc.nasa.gov/amon_nu_em_coinc_events.html}. The event characteristics and the corresponding upper limits are presented in Table~\ref{tab:NuEM_publis} and Table~\ref{tab:NuEM_limits}, together with the reference of the GCN circulars where the results have been published.\\

\begin{table}[h!]{
 \begin{center}{
\begin{tabular}{|c|c|c|c|c|c|c|}
\hline
Alert & Trigger & RA,Dec & R$_{err}$ &  Duration & Visibility   & GCN\\
         & Id	  & (deg)    & (deg)& (sec)	  & (\%)	& Id\\
\hline
NuEM-220116A & 0\_105322 & 322.13, $+$27.26 & 0.17 & 23130.4   & 34   & 31476 \\
NuEM-211209A & 0\_101674  & 12.03, $-$5.75     & 0.18 & 18273.3   & 53   & 31198  \\
NuEM-211020A & 0\_96720   &  99.76, $+$9.07   & 0.17 & 21670.1    & 45   & 30954  \\
NuEM-210515B  & 0\_85791   &  93.93, $+$12.51 & 0.2 & 22165.2     & 42  & 30024  \\
NuEM-210515A  & 0\_85790  &  93.64, $+$14.66 & 0.15 & 22443     & 42   & 30024  \\
NuEM-210111A  & 0\_73310    &  162.34, $-$19.46 & 0.37 & 22742.5 & 39   & 29294  \\
NuEM-201124A & 0\_68186    &  135.0, $+$7.74    & 0.23 & 21531.2   & 46   & 28953  \\
NuEM-201107A & 0\_66291    &  140.2, $+$29.76  & 0.15 & 23105.4   & 30   &  /  \\
NuEM-200717A & 0\_54519    &  118.5, $-$1.62      & 0.38 & 19395.5   & 50   & 28144  \\
  \hline
\end{tabular}}
\caption{Summary of the ANTARES follow-up of the NuEM triggers (IceCube / HAWC coincidence). R$_{\rm err}$ is the 50~\% allowed region. "Duration" corresponds to the duration of the coincidence. The last column gives the GCN circular references of the ANTARES follow-up.}
\label{tab:NuEM_publis}
\end{center}}
\end{table}

\begin{table}{
\begin{center}{
\begin{tabular}{|c|c|c|c|c|c|}
\hline
Alert & \multicolumn{2}{c|}{Fluence U.L. (GeV cm$^{-2}$) at 90~\% C.L.} \\
             &  $dN/dE\propto E^{-2}$& $dN/dE\propto E^{-2.5}$ \\
\hline
NuEM-220116A & 14 (6 TeV - 6 PeV)  &  31 (1 - 580 TeV)  \\
NuEM-211209A & 15 (3 TeV - 3.3 PeV)  &  26 (0.5 - 290 TeV)  \\
NuEM-211020A & 15 (4 TeV - 4 PeV)  &  26 (0.75 - 380 TeV)  \\
NuEM-210515B & 15 (5 TeV - 4 PeV)  &  30 (1 - 400 TeV)  \\
NuEM-210515A & 15 (5 TeV - 4 PeV)  &  30 (1 - 400 TeV)  \\
NuEM-210111A & 14 (5 TeV - 5 PeV)  &  35 (1 - 480 TeV)  \\
NuEM-201124A & 19 (4 TeV - 4 PeV)  &  31 (0.72 - 380 TeV)  \\
NuEM-200717A & 15 (3 TeV - 3.4 PeV)  &  25 (0.6 - 310 TeV)  \\
\hline
\end{tabular}}
\caption{Upper limits (at 90~\% C.L.) on neutrino fluence as derived from the non-observation of ANTARES coincidences for each NuEM triggers. For each upper limit, the energy range in which 90~\% of events are observed (excluding the 5~\% of events with the lower/higher energies) is given.}
\label{tab:NuEM_limits}
\end{center}}
\end{table}

%% file: conclusion.tex
As a coincident observation by two experiments significantly decreases the probability of false alerts, fast confirmation is essential to allow observatories with limited follow-up capabilities, e.g., due to limited sky coverage or observation time, to efficiently prioritise and schedule their resources. A further advantage is that a subsequent offline analysis of data collected by different instruments upon an alert may yield a statistically relevant result from a combination of signals that by themselves would not be considered significant enough \color{blue}to report \color{black}.\\

Public alerts are common for EM transients, especially gamma-ray bursts, soft-gamma repeaters, supernovae, etc. To study the parameters of physical processes inherent to these astrophysical sources, it is necessary to collect as much as possible wide multi-wavelength and multi-probe information as possible. This can only happen with a synergy between different instruments based on efficient, fast and reliable communication between them. The GCN is in the centre of this strategy as a fast dispatcher of triggers and results. Recently, multi-messenger actors have also adopted a strategy similar to public alert distribution: IceCube in 2016, LIGO/Virgo in 2019, HAWC in 2020. With more than one hundred triggers in one year from the \color{blue} O3 run \color{black} of LIGO/Virgo and the new alert selection of IceCube, some maturity has been reached. \color{blue} A fully automatised online analysis framework has been implemented in the ANTARES Collaboration \color{black}, that looks for time/space coincidences with the time/direction of EM, neutrino and GW  transient external triggers. All the received public alerts have been followed provided that at the time of the trigger, their position in the sky was below the horizon for the ANTARES detector. Despite the fact that no coincidences have been found in the online analysis, this effort has highlighted the multi-messenger program of the ANTARES neutrino telescope to a broad community. Most of the remaining work consists in the writing and validation of each GCN circular. This step can also be automatised in the future. \\

KM3NeT~\cite{KM3Net:2016zxf} is starting to take data with a sensitivity larger than ANTARES, and this new detector will allow multi-flavor neutrino detection in real time with an unprecedented angular resolution~\cite{Assal:2021lbu}. For the muon-neutrino golden channel, the angular precision can be as low as 0.1$^{\circ}$ at very high energies. In KM3NeT, all-flavor neutrino events will be used for online follow-up studies in a large energy range from a few GeV to a few PeV.\\

%% file: main.bbl
\begin{thebibliography}{10}

\bibitem{Collaboration:2011nsa}
M.~Ageron {\em et~al.}, ``{ANTARES: the first undersea neutrino telescope},''
  {\em Nucl. Instrum. Meth. A}, vol.~656, pp.~11--38, 2011.

\bibitem{KM3Net:2016zxf}
S.~Adrian-Martinez {\em et~al.}, ``{Letter of intent for KM3NeT 2.0},'' {\em J.
  Phys. G}, vol.~43, no.~8, p.~084001, 2016.

\bibitem{2016JCAP...02..062A}
S.~Adri\'an-Mart\'\i{}nez {\em et~al.}, ``{Optical and X-ray early follow-up of
  ANTARES neutrino alerts},'' {\em JCAP}, vol.~02, p.~062, 2016.

\bibitem{2021ApJ...911...48A}
A.~Albert {\em et~al.}, ``{ANTARES Search for Point Sources of Neutrinos Using
  Astrophysical Catalogs: A Likelihood Analysis},'' {\em Astrophys. J.},
  vol.~911, no.~1, p.~48, 2021.

\bibitem{Ageron:2011pe}
M.~Ageron {\em et~al.}, ``{The ANTARES Telescope Neutrino Alert System},'' {\em
  Astropart. Phys.}, vol.~35, pp.~530--536, 2012.

\bibitem{2018arXiv180508675A}
A.~Albert {\em et~al.}, ``{Long-term monitoring of the ANTARES optical module
  efficiencies using$^{40}\mathrm{{K}}$ decays in sea water},'' {\em Eur. Phys.
  J. C}, vol.~78, no.~8, p.~669, 2018.

\bibitem{ANTARES:2017bia}
A.~Albert {\em et~al.}, ``{Search for High-energy Neutrinos from Binary Neutron
  Star Merger GW170817 with ANTARES, IceCube, and the Pierre Auger
  Observatory},'' {\em Astrophys. J. Lett.}, vol.~850, no.~2, p.~L35, 2017.

\bibitem{Molla:2020wsw}
A.~Albert {\em et~al.}, ``{Search for neutrino counterparts of
  gravitational-wave events detected by LIGO and Virgo during run O2 with the
  ANTARES telescope},'' {\em Eur. Phys. J. C}, vol.~80, no.~5, p.~487, 2020.

\bibitem{Adrian-Martinez:2016xij}
S.~Adrián-Martínez {\em et~al.}, ``{Stacked search for time shifted high
  energy neutrinos from gamma ray bursts with the ANTARES neutrino
  telescope},'' {\em Eur. Phys. J. C}, vol.~77, no.~1, p.~20, 2017.

\bibitem{2021ApJ...910....4A}
R.~Abbasi {\em et~al.}, ``{Follow-up of Astrophysical Transients in Real Time
  with the IceCube Neutrino Observatory},'' {\em Astrophys. J.}, vol.~910,
  no.~1, p.~4, 2021.

\bibitem{Smith:2012eu}
M.~Smith {\em et~al.}, ``{The Astrophysical Multimessenger Observatory Network
  (AMON)},'' {\em Astropart. Phys.}, vol.~45, pp.~56--70, 2013.

\bibitem{2018ApJ...863L..30A}
A.~Albert {\em et~al.}, ``{The Search for Neutrinos from TXS 0506+056 with the
  ANTARES Telescope},'' {\em Astrophys. J. Lett.}, vol.~863, no.~2, p.~L30,
  2018.

\bibitem{2021ApJ...920...50A}
A.~Albert {\em et~al.}, ``{Search for Neutrinos from the Tidal Disruption
  Events AT2019dsg and AT2019fdr with the ANTARES Telescope},'' {\em Astrophys.
  J.}, vol.~920, no.~1, p.~50, 2021.

\bibitem{2019ApJ...879..108A}
A.~Albert {\em et~al.}, ``{ANTARES neutrino search for time and space
  correlations with IceCube high-energy neutrino events},'' {\em Astrophys.
  J.}, vol.~879, no.~2, p.~108, 2019.

\bibitem{Meszaros:2006rc}
P.~Meszaros, ``{Gamma-Ray Bursts},'' {\em Rept. Prog. Phys.}, vol.~69,
  pp.~2259--2322, 2006.

\bibitem{2016PhRvD.93l3011M}
R.~Moharana, S.~Razzaque, N.~Gupta, and P.~Meszaros, ``{High Energy Neutrinos
  from the Gravitational Wave event GW150914 possibly associated with a short
  Gamma-Ray Burst},'' {\em Phys. Rev. D}, vol.~93, no.~12, p.~123011, 2016.

\bibitem{Kotera:2016dmp}
K.~Kotera and J.~Silk, ``{Ultrahigh Energy Cosmic Rays and Black Hole
  Mergers},'' {\em Astrophys. J. Lett.}, vol.~823, no.~2, p.~L29, 2016.

\bibitem{Perna:2016jqh}
R.~Perna, D.~Lazzati, and B.~Giacomazzo, ``{Short Gamma-Ray Bursts from the
  Merger of Two Black Holes},'' {\em Astrophys. J. Lett.}, vol.~821, no.~1,
  p.~L18, 2016.

\bibitem{Bartos:2016dgn}
I.~Bartos, B.~Kocsis, Z.~Haiman, and S.~Márka, ``{Rapid and Bright
  Stellar-mass Binary Black Hole Mergers in Active Galactic Nuclei},'' {\em
  Astrophys. J.}, vol.~835, no.~2, p.~165, 2017.

\bibitem{2017MNRAS.464..946S}
N.~C. Stone, B.~D. Metzger, and Z.~Haiman, ``{Assisted inspirals of stellar
  mass black holes embedded in AGN discs: solving the \textquoteleft{}final au
  problem\textquoteright{}},'' {\em Mon. Not. Roy. Astron. Soc.}, vol.~464,
  no.~1, pp.~946--954, 2017.

\bibitem{2016ApJ...822L...9M}
K.~Murase, K.~Kashiyama, P.~M\'esz\'aros, I.~Shoemaker, and N.~Senno,
  ``{Ultrafast Outflows from Black Hole Mergers with a Minidisk},'' {\em
  Astrophys. J. Lett.}, vol.~822, no.~1, p.~L9, 2016.

\bibitem{PhysRevD.98.043020}
S.~S. Kimura, K.~Murase, I.~Bartos, K.~Ioka, I.~S. Heng, and P.~M\'esz\'aros,
  ``{Transejecta high-energy neutrino emission from binary neutron star
  mergers},'' {\em Phys. Rev. D}, vol.~98, no.~4, p.~043020, 2018.

\bibitem{2017ApJ848L4K}
S.~S. Kimura, K.~Murase, P.~M\'esz\'aros, and K.~Kiuchi, ``{High-Energy
  Neutrino Emission from Short Gamma-Ray Bursts: Prospects for Coincident
  Detection with Gravitational Waves},'' {\em Astrophys. J. Lett.}, vol.~848,
  no.~1, p.~L4, 2017.

\bibitem{Kimura:2019ipr}
S.~S. Kimura, ``{High-energy emissions from neutron star mergers},'' {\em EPJ
  Web Conf.}, vol.~210, p.~03001, 2019.

\bibitem{Fang:2017tla}
K.~Fang and B.~D. Metzger, ``{High-Energy Neutrinos from Millisecond Magnetars
  formed from the Merger of Binary Neutron Stars},'' {\em Astrophys. J.},
  vol.~849, no.~2, p.~153, 2017.

\bibitem{Abbott:2016blz}
B.~Abbott {\em et~al.}, ``{Observation of Gravitational Waves from a Binary
  Black Hole Merger},'' {\em Phys. Rev. Lett.}, vol.~116, no.~6, p.~061102,
  2016.

\bibitem{Adrian-Martinez:2016xgn}
S.~Adrian-Martinez {\em et~al.}, ``{High-energy Neutrino follow-up search of
  Gravitational Wave Event GW150914 with ANTARES and IceCube},'' {\em Phys.
  Rev. D}, vol.~93, no.~12, p.~122010, 2016.

\bibitem{ANTARES:2017iky}
A.~Albert {\em et~al.}, ``{Search for High-energy Neutrinos from Gravitational
  Wave Event GW151226 and Candidate LVT151012 with ANTARES and IceCube},'' {\em
  Phys. Rev. D}, vol.~96, no.~2, p.~022005, 2017.

\bibitem{GBM:2017lvd}
B.~Abbott {\em et~al.}, ``{Multi-messenger Observations of a Binary Neutron
  Star Merger},'' {\em Astrophys. J. Lett.}, vol.~848, no.~2, p.~L12, 2017.

\bibitem{Baret:2011tk}
B.~Baret {\em et~al.}, ``{Bounding the Time Delay between High-energy Neutrinos
  and Gravitational-wave Transients from Gamma-ray Bursts},'' {\em Astropart.
  Phys.}, vol.~35, pp.~1--7, 2011.

\bibitem{Abbott:2020uma}
B.~Abbott {\em et~al.}, ``{GW190425: Observation of a Compact Binary
  Coalescence with Total Mass $\sim$3.4 M$_{\odot}$.},'' {\em Astrophys. J.
  Lett.}, vol.~892, p.~L3, 2020.

\bibitem{LIGOScientific:2020stg}
R.~Abbott {\em et~al.}, ``{GW190412: Observation of a Binary-Black-Hole
  Coalescence with Asymmetric Masses},'' {\em Phys. Rev. D}, vol.~102, no.~4,
  p.~043015, 2020.

\bibitem{ColomerMolla:2020dln}
M.~Colomer~Molla, B.~Baret, A.~Coleiro, D.~Dornic, and T.~Pradier, ``{Search
  for neutrino counterparts of cataloged gravitational-wave events detected by
  Advanced-LIGO and Virgo during run O2 with ANTARES},'' {\em PoS},
  vol.~ICRC2019, p.~856, 2020.

\bibitem{2017PhRvD..96h2001A}
A.~Albert {\em et~al.}, ``{First all-flavor neutrino pointlike source search
  with the ANTARES neutrino telescope},'' {\em Phys. Rev. D}, vol.~96, no.~8,
  p.~082001, 2017.

\bibitem{Albert:2016eyr}
A.~Albert {\em et~al.}, ``{The search for high-energy neutrinos coincident with
  fast radio bursts with the ANTARES neutrino telescope},'' {\em Mon. Not. Roy.
  Astron. Soc.}, vol.~482, no.~1, pp.~184--193, 2019.

\bibitem{2021MNRAS.500.5614A}
A.~Albert {\em et~al.}, ``{Constraining the contribution of Gamma-Ray Bursts to
  the high-energy diffuse neutrino flux with 10 yr of ANTARES data},'' {\em
  Mon. Not. Roy. Astron. Soc.}, vol.~500, no.~4, pp.~5614--5628, 2020.

\bibitem{Assal:2021lbu}
W.~Assal, D.~Dornic, F.~Huang, E.~Le~Guirriec, M.~Lincetto, and G.~Vannoye,
  ``{Real-time multi-messenger analysis framework for KM3NeT},'' {\em JINST},
  vol.~16, no.~09, p.~C09034, 2021.

\end{thebibliography}
